\documentclass[eqsecnum,showpacs,aps]{revtex4}
\usepackage{graphicx}

\newcommand{\ac}{ {\alpha_{c}} }

\newcommand{\muu}{ {\mu_{u}} }
\newcommand{\mud}{ {\mu_{d}} }
\newcommand{\mus}{ {\mu_{s}} }
\newcommand{\mue}{ {\mu_{e}} }

\newcommand{\simg}
    {\mathrel{\raise.3ex\hbox{$>$\kern-.75em\lower1ex\hbox{$\sim$}}}}
\newcommand{\siml}
    {\mathrel{\raise.3ex\hbox{$<$\kern-.75em\lower1ex\hbox{$\sim$}}}}

\preprint{OU-TAP 220}

\begin{document}

\draft


\title{Restricting quark matter models by gravitational wave observation}

\author{
Hajime Sotani$^{1}$
\footnote{Electronic address:sotani@gravity.phys.waseda.ac.jp}}
\author{
Kazunori Kohri$^{2}$
\footnote{Electronic address:kohri@vega.ess.sci.osaka-u.ac.jp}}
\author{
Tomohiro Harada$^{3}$
\footnote{Electronic address:T.Harada@qmul.ac.uk}}

\address{
$^{1}$ Department of Physics, Waseda University,
Okubo 3-4-1, Shinjuku, Tokyo 169-8555, Japan
}

\address{
$^{2}$ Department of Earth and Space Science,
Graduate School of Science, Osaka University,
Toyonaka, Osaka 560-0043, Japan
}

\address{
$^{3}$ Astronomy Unit, School of Mathematical Sciences, Queen Mary, 
 University of London, Mile End Road, London E1 4NS, UK
}

\date{\today}

\begin{abstract}
    We consider the possibilities for obtaining information about the equation
    of state for quark matter by using future direct observational
    data on gravitational waves.  We study the nonradial oscillations
    of both fluid and spacetime modes of pure quark stars.  If
    we observe the $f$ and the lowest $w_{\rm II}$ modes from quark
    stars, by using the simultaneously obtained radiation radius we
    can constrain the bag constant $B$ with reasonable accuracy,
    independently of the $s$ quark mass.
\end{abstract}




\maketitle


\section{Introduction}

Towards observing gravitational waves,  several
gravitational wave interferometers on Earth are steadily
developing today, such as the Laser Interferometric Gravitational Wave
Observatory (LIGO) \cite{Althouse1992}, TAMA300 \cite{Tsubono1995},
GEO600 \cite{Hough1996}, and VIRGO \cite{Giazotto1990}.  Thus it will
be possible for us to detect gravitational waves directly in the near
future.
It is believed that mergers of binary neutron-star-neutron-star (NS-NS),
neutron-star-black-hole (NS-BH), and BH-BH,
or supernovae, and so on, can  become strong sources of
gravitational waves.  After these violent events occur, compact objects
may be left and may be turbulent. Then gravitational waves are emitted from them.
At this time, these gravitational waves convey information on the source
object.  If these gravitational waves are directly detected  on
Earth, it is possible to obtain some information about the
sources. This research field is called ``gravitational wave
astronomy.''  In this field, there is an attempt to obtain information
about properties of the equation of state (EOS) of high density
matter. This is one of the most important purposes of gravitational
wave astronomy.

Gravitational waves are emitted by the nonspherical oscillation of compact
objects. The oscillations are damped out as gravitational waves carry
away the oscillational energy.  Such oscillations are called
quasinormal modes (QNMs). The QNMs have complex frequencies whose real
and imaginary parts correspond to the oscillational frequency and damping
rate, respectively.  The QNMs fall into two types from their nature.
One involves fluid modes which are connected with the stellar matter. The
other involves spacetime modes which are the oscillations of spacetime
metric.  Moreover, the fluid modes are classified into various
types. The well-known modes are the $f$, $p$, and $g$
modes \cite{Kokkotas1999}. The fluid modes have a characteristic  that
the damping rate Im($\omega$) is much smaller than the oscillational
frequency Re($\omega$).  The
$f$ mode is the fundamental mode. There exists only one $f$ mode for
each index $l$ of spherical harmonics $Y_{lm}$.  The $p$ mode is the
pressure or acoustic mode, whose restoring force is caused by the
pressure gradient inside the star.  The $g$ mode is the gravity mode,
which arises from buoyancy in a gravity field.  The $w$ and $w_{\rm
II}$ modes are spacetime
modes \cite{Kokkotas1992,Leins1993}. Unlike the fluid modes, the
damping rate of the $w$ and $w_{\rm II}$ modes is comparable to or
larger than the oscillational frequency.

For the QNMs of neutron stars, so far many authors have argued
the possibilities for determining the EOS in the high density region and/or
for restricting the properties of neutron stars, such as the radius $R$ or
mass $M$, by employing the observed gravitational wave of  several
nonradial modes \cite{Andersson1996,Andersson1998,Kokkotas1999,
KokkotasAndersson2001,Kokkotas2001,Benhar2000,Lindblom:1992,harada:2001}.
As a candidate for a star which is smaller than neutron stars,
the possibility of a quark star or a compact star, which is supported by
degenerate pressure of quark matter, has been pointed out.  Such a
quark star has been investigated by many authors (see, e.g.,
Refs. \cite{ivanenko:1969,itoh:1970,Collins:1974ky,Baym:yu,Chapline:gy,
Kislinger:1978,Fechner:ji,Freedman:1977gz,Baluni:1977mk,Witten:1984rs,
Alcock:1986hz,Haensel:qb,Rosenhauer,Glendenning,Schertler,
Zdunik,Sinha,Gerlach} and references therein).  In their view, it
is commonly assumed that such quark stars contain quark matter in the
core region and are surrounded by hadronic matter, although they are in
the branch of neutron stars \cite{Rosenhauer}. Witten
\cite{Witten:1984rs} suggested another type of quark star. If the
true ground state of hadrons is bulk quark matter, which consists of
approximately equal numbers of $u$, $d$, and $s$ quarks (``strange
matter''), there exist self-bound quark stars. They are called
``strange stars.'' In this case, their mass and radius are smaller
than those of typical neutron stars, which are $\sim$ 10 km and
$\sim1.4M_{\odot}$, respectively.  Because we do not have reliable
information about the equilibrium properties of hadronic and quark
matters at high densities, it is not clear what kinds of quark stars are
realized.

Recently, Drake $et$ $al$. reported that the deep Chandra LETG+HRC-S
observations of the soft X-ray source RX J1856.5--3754 reveal an X-ray
spectrum quite close to that of a blackbody of temperature $T= 61.2
\pm 1.0$ eV \cite{Drake:2002bj}.  The data contain evidence for
the lack of spectral features or pulsation \cite{footnote}.  Drake $et$
$al$. also reported that the interstellar medium neutral hydrogen column
density is $N_H=(0.8$--$1.1)\times 10^{20}$ cm$^{-2}$.  With the
results of recent HST parallax analyses, that yields an estimate of
111--170 pc for distance $D$ to RX J1856.5-3754. Combining this
range of $D$ with the blackbody fit leads to a radiation radius of
$R_{\infty}=3.8$--8.2 km. That is smaller than typical neutron star
radii \cite{Lattimer:2001}.  Thus they suggested that the X-ray source
may be a quark star.

In the meanwhile,  in Ref. \cite{Walter:2002}, Walter and Lattimer
claimed that the blackbody model adopted in Ref. \cite{Drake:2002bj}
could not explain the observed UV-optical spectrum. They undertook to
fit the two-temperature blackbody and heavy-element atmosphere models
which were discussed in Ref. \cite{Pons:2001px} in X-ray and
UV-optical wavelengths. From their analyses they found that the
radiation radius was 12--26 km and was consistent with that of a
neutron star. However, this model cannot explain the lack of spectral
features.  In addition, recently Braje and Romani also suggested a
two-temperature blackbody model, which can reproduce both the X-ray
and optical-UV spectral \cite{Braje:2002}. Although their model is
inconsistent with the fact that pulsation is not detected, this might be
explained  by the object being a young normal pulsar and its
nonthermal radio beam missing the Earth's line of sight.  However, they can
not also answer why there are no features in the observed X-ray
spectrum. In Ref. \cite{Burwitz:2002vm}, Burwitz $et$ $al$. discussed the
possibility of a condensate surface which is made of unknown material to
explain both the UV-optical and X-ray spectra in a neutron star.
More recently, some groups discussed the possibility that the effects
of a strong magnetic field ($B\sim 10^{13}\ \mbox{G}$) \cite{Thoma2003} or
rapid rotation ($P<10\ \mbox{ms}$) \cite{Pavlov2003} may smear out any
spectral features.  In these situations, however, it seems that there
are no reliable models that account for all the observational
facts. It is still controversial whether RXJ1856.5-3754 is a normal
neutron star or some other compact star like a quark star.
Here we adopt the simplest picture, which is
used by Drake $et$ $al$. \cite{Drake:2002bj}:  the uniform temperature
blackbody model.

As for gravitational waves emitted from quark stars, Yip, Chu, and
Leung studied nonradial stellar oscillation for stars whose radius is
around 10 km~\cite{Yip1999}.  Kojima and Sakata demonstrated the
possibility of distinguishing quark stars from neutron stars by using
both the oscillational frequency and the damping rate of the $f$
mode~\cite{Kojima2002}.  Sotani and Harada showed that the $f$ and
lowest $w_{\rm II}$ modes depend strongly on the EOS of quark matter
and the properties of quark stars, where the lowest $w_{\rm II}$ mode
is the one which has the largest frequency among all $w_{\rm II}$
modes~\cite{Sotani2003}.  Then they pointed out the possibility of
determining the EOS and/or the stellar properties.
Furthermore, they also studied $w$ modes in detail. However, they
clearly showed that $w$ modes do not depend much on the EOS of quark
matter and are not important for constraining the model parameters
from observations.
In their work, however, they assumed that  the star is a pure quark
star and that the EOS is described by a simple bag model which has
only one parameter: i.e., the bag constant $B$. In general, there are
a variety of parameters even within bag models: e.g., the bag constant
$B$, the strange quark mass $m_s$,  the fine structure constant in QCD
$\ac$, and so on.  In particular, if the effects of nonvanishing
strange quark mass $m_{s}$ are taken into account, the structure of
quark stars can be affected considerably (for recent analyses, see
Ref.~\cite{Kohri:2002hf} and references therein). In this situation,
we compute the QNMs in the bag model used in Ref.~cite{Kohri:2002hf}
and investigate the possibility of restricting  the model parameters
by the observations of QNMs. In this study, we deal with only $f$  and
the lowest $w_{\rm II}$ modes
in response to the results in Ref.~\cite{Sotani2003}.

Effective theories of quantum chromodynamics (QCD), such as bag models,
perturbation theories, or finite-temperature lattice data, are difficult
to test by experiments, especially in a low temperature and
high density regime.
To further study them and fit their model
parameters, we should compare the theoretical predictions in such models
with experimental data---e.g., data in relativistic heavy ion collision
experiments and so on.
Thus, in this situation information about compact objects
obtained by astrophysical observations
is indispensably valuable, being independent
of the above-mentioned ground-based
experiments in nuclear physics or particle physics.
Among the astrophysical observations,
the observation of gravitational waves emitted
from oscillating compact objects is quite unique
because gravitational waves directly convey information on
the internal structure of compact objects,
where the density would reach the nuclear density.

The plan of this paper is as follows.  In Sec. \ref{sec:QuarkStar} we
introduce the basic equation including the EOS to construct quark stars
as the source of gravitational waves and to give the properties of quark
star structure. We present a method of determining the QNMs
for the case of spherically symmetric stars in
Sec. \ref{sec:method}. In Sec. \ref{sec:result} we show the  numerical
results for the QNMs for the quark star
constructed in Sec. \ref{sec:QuarkStar}. In this section we present
the dependence of the QNMs on the parameters of the EOS and
stellar properties, and discuss the possibility of determining the EOS of
quark matter.  We conclude this paper in Sec. \ref{sec:conclusion}.  We
adopt units of $c=\hbar=G=1$, where $c$, $\hbar$,
and $G$ denote the speed of light, reduced Planck's constant, and
gravitational constant, respectively, and the metric signature of
$(-,+,+,+)$ throughout this paper.

\section{Quark Star Models}
\label{sec:QuarkStar}

We assume that the quark star is static and
spherically symmetric.  In this case  the metric is described by
\begin{equation}
    ds^{2}=-e^{2\Phi}dt^{2}+e^{2\Lambda}dr^{2}+r^{2}\left(d\theta^{2}+\sin^{2}
             \theta d\phi^{2}\right),
    \label{metric_star}
\end{equation}
where $\Phi$ and $\Lambda$ are metric functions of $r$ and $\Lambda$ is
related to the mass function $m(r)$ as
\begin{equation}
      m(r) = \frac{1}{2}r\left(1-e^{-2\Lambda}\right). \label{ephi}
\end{equation}
The mass function $m(r)$ is the gravitational mass inside a surface
of radius $r$ and satisfies
\begin{equation}
     \frac{dm}{dr}    = 4\pi r^2\rho, \label{dmdr}
\end{equation}
where $\rho$ is the energy density. The equilibrium stellar model is
constructed
by solving the equations
\begin{eqnarray}
      \frac{dP}{dr}   &=& -\frac{\left(\rho+P\right)\left(m+4\pi r^3P\right)}
                           {r\left(r-2m\right)}, \label{dpdr} \\
      \frac{d\Phi}{dr}   &=& \frac{\left(m+4\pi r^3P\right)}
                           {r\left(r-2m\right)}. \label{dphidr}
\end{eqnarray}
Equation (\ref{dpdr}) is the Tolman-Oppenheimer-Volkoff (TOV)
equation \cite{Oppenheimer:1939ne}.  In addition to the above
equations we need an EOS to compute the star configuration. The
stellar surface $r=R$ is the position where the pressure vanishes, and
the stellar mass $M$ is defined as $M=m(R)$.  Because the metric in
the exterior of the star is the Schwarzschild one, the metric
(\ref{metric_star}) must be connected smoothly with the Schwarzschild one
at the stellar surface.

Next we describe the EOS in bag models.  To construct stars composed of
zero-temperature $uds$ quark matter, we  start with the thermodynamic
potentials $\Omega_q$ for a homogeneous gas of  $q$ quarks
($q=u,d,s$) of rest mass $m_q$ up to first order in the QCD coupling
constant $\ac$.  The expressions for $\Omega_q$ are given as a sum of
the kinetic  term and the one-gluon-exchange term at the
renormalization scale
$\Lambda=m_q$ \cite{Baym:1975va,Freedman:1976xs,Baluni:1977mk}.  Here
we assume that $m_u=m_d=0$. Then $\Omega_q$ is expressed by
\begin{eqnarray}
     \label{eq:Omegau}
     \Omega_{u} &=& -
     \frac{\muu^{4}}{4\pi^{2}}\left(1-\frac{2\ac}{\pi}\right), \\
     \label{eq:Omegad}
     \Omega_{d} &=&
     - \frac{\mud^{4}}{4\pi^{2}}\left(1-\frac{2\ac}{\pi}\right), \\
     \label{eq:Omegas}
     \Omega_{s} &=&
     - \frac{m_s^{4}}{4\pi^{2}}\left\{
       x_s\eta_s^3-\frac32F(x_s) -\frac{2\ac}{\pi}\left[ 
           3F(x_s)\left(F(x_s)+2\ln{x_s} \right) -2\eta_s^{4}
           +6\left(\ln{\frac{\Lambda}{\mus}}\right)
           F(x_s) \right]
     \right\}, 
\end{eqnarray}
with $F(x_s) = x_s\eta_s-\ln{(x_s+\eta_s)}$,  where $x_{s} =
\mus/m_{s}$, $\eta_{s}=\sqrt{x_{s}^{2}-1}$, $\mu_{q}$ is the  chemical
potential of $q$ quarks, $\ac = \ac(\Lambda)$, and $m_{s} =
m_{s}(\Lambda)$.  Here we choose $\Lambda = \mu_s$.
The electron thermodynamic potential $\Omega_e$ in the massless
noninteracting form is given by
\begin{eqnarray}
         \label{eq:Omegae}
     \Omega_{e} &=& - \frac{\mue^{4}}{12\pi^{2}},
\end{eqnarray}
where $\mue$ is the electron chemical potential.

Within the bag model, we express the total energy density $\rho$
as
\begin{eqnarray}
     \label{eq:energy}
     \rho = \sum_{i=u,d,s,e}\left(\Omega_{i} +  \mu_{i}n_{i}\right) + B,
\end{eqnarray}
where $B$ is the bag constant \cite{Farhi:1984qu} which is the excess
energy density effectively representing the nonperturbative color
confining interactions. $n_i$ is the number density of $i$ particles
given by $n_{i} = - {\partial\Omega_{i}}/{\partial\mu_{i}}$.  Then,
the pressure $p$ is represented by
\begin{eqnarray}
     \label{eq:pressure}
     p &=& -    \sum_{i=u,d,s,e}\Omega_{i} - B.
\end{eqnarray}

The parameters $m_s$, $\ac$, and $B$ were obtained by fits to
light-hadron spectra; e.g., see
Refs. \cite{DeGrand:cf,Carlson:er,Bartelski:1983cc} (also see
Table \ref{table:fitting}).  Note that these parameters  do not
correspond directly to the values appropriate to  bulk quark matter
\cite{Farhi:1984qu}.  When we allow for possible uncertainties, we set
$145\leq B^{1/4}\leq 200$ MeV, $0\leq m_s \leq 300$ MeV, and
$0\leq\ac\leq0.9$.  The ranges of $m_s$ and $\ac$ are not always
consistent with the values in Table \ref{table:fitting}.  However,
when we renormalize them at an energy scale of interest to us, they are
consistent with values of the Particle Data Group \cite{PDG}.

To obtain the equilibrium composition of the ground-state matter at a
given baryon density or chemical potential, the conditions for
equilibrium with respect to the weak interaction and for overall
charge neutrality are required.  These conditions are expressed by
\begin{eqnarray}
     \label{eq:weak-equiv1}
     \muu &=& \mud - \mue, \\
     \label{eq:weak-equiv2}
     \mud &=&  \mus,
\end{eqnarray}
and
\begin{eqnarray}
     \label{eq:charge_neutral}
     \frac23n_{u} - \frac13n_{d} - \frac13 n_{s} =  n_{e}.
\end{eqnarray}
{}From Eqs. (\ref{eq:Omegau})--(\ref{eq:charge_neutral}), we can
obtain the energy density and pressure as a function of $n_{B}$
(or $\mu_B$). In Ref. \cite{Sotani2003} the authors
assumed that $m_{s}=0$ and $\alpha_{c}=0$. Then, the EOS is
analytically given  by
\begin{equation}
    P=\frac{1}{3}(\rho-4B). \label{bagmodel}
\end{equation}
In this case, the density at the stellar surface is given by $\rho(R)=4B$.

\begin{center}
\begin{figure}[htbp]
\includegraphics{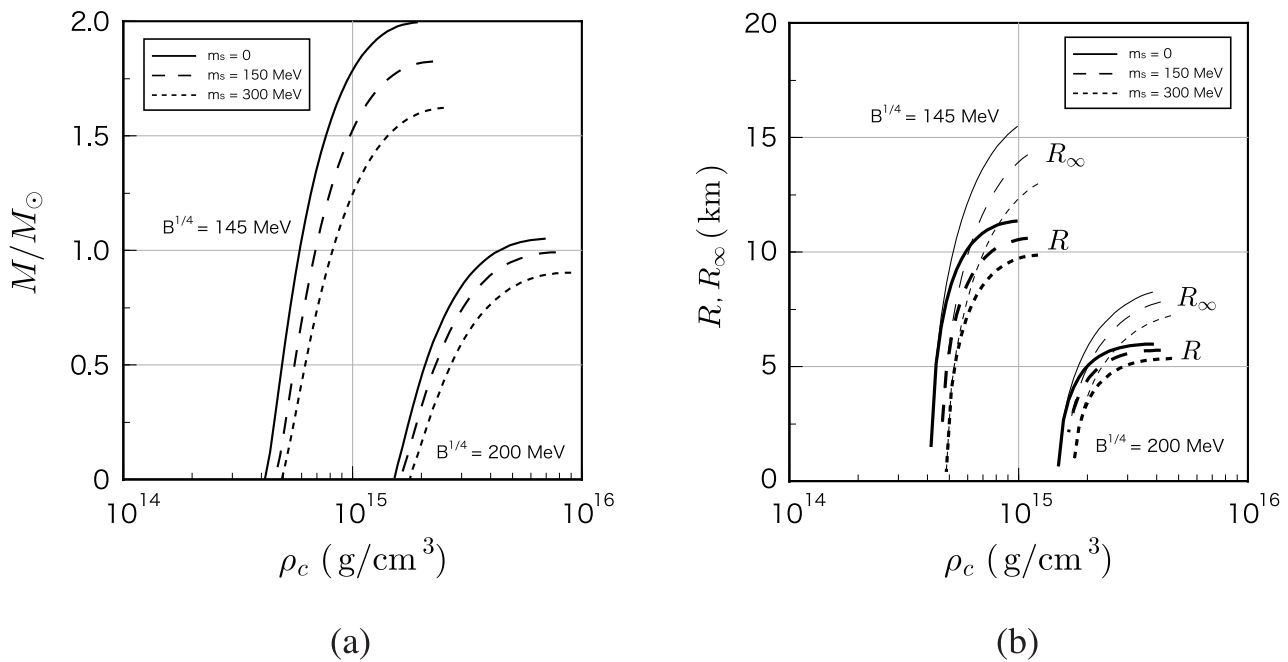}
\caption{
(a) Mass of the star as a function of central density and (b) radius
of the star  as a function of central density (thick lines) for
$B^{1/4}=145$ MeV and $200$ MeV.  Thin lines denote the radiation
radius $R_{\infty}$. Here we adopt three cases of the $s$ quark
mass: i.e., $m_s=0$ (solid line), $150$ MeV (dashed line), and $300$
MeV (dotted line).  We adopt $\alpha_c = 0.6$
as a value of the fine structure constant in QCD.
}
\label{fig_property}
\end{figure}
\end{center}

In Fig. \ref{fig_property}, we plot several sequences of quark star
models with different values of the bag constant $B$ and the $s$ quark
mass $m_s$.  Here we have fixed the fine structure constant in QCD as
$\alpha_c=0.6$ because the dependence of properties of quark stars on
$\alpha_c$ is weak, as indicated in Ref. \cite{Kohri:2002hf}.
Therefore we deal with stellar models only for the case of
$\alpha_c=0.6$ in this paper.  In Fig. \ref{fig_property}(a) we plot
the mass $M/M_{\odot}$ of the star as a function of central density
$\rho_c$. Here we set $B^{1/4}=145$, $200$ MeV and $m_s=0$, $150$,
$300$ MeV, respectively.  In Fig. \ref{fig_property}(b) we also plot
the radius of the star  as a function of central density by thick
lines.  Thin lines denote the radiation radius $R_{\infty}$. In
Fig. \ref{fig_property}, the solid line, dashed line, and dotted line
correspond to the cases of $m_s=0$, $150$, and $300$ MeV, respectively.
In Tables \ref{tab_property_QS1} and \ref{tab_property_QS2}, we list
several properties of quark stars whose radiation radii are within the
range of $R_{\infty}=3.8$--$8.2$ km, which was reported by
Ref. \cite{Drake:2002bj}.

\section{Determination of the QNM}
\label{sec:method}

Since we are interested in the dependence of QNMs on the EOS of quark
matter, we consider only polar perturbation.  If we adopt the
Regge-Wheeler gauge, the metric perturbation is given by
     \begin{equation}
      g_{\mu\nu} = g^{(B)}_{\mu\nu} + \left(
      \begin{array}{cccc}
      r^l\hat{H}e^{2\Phi} & i\omega r^{l+1}\hat{H}_{1} & 0 & 0  \\
      i\omega r^{l+1}\hat{H}_{1} & r^l\hat{H}e^{2\Lambda} & 0 & 0  \\
      0 & 0 & r^{l+2}\hat{K} & 0 \\
      0 & 0 & 0              & r^{l+2}\hat{K}\sin^{2}\theta
     \end{array}
     \right) Y^{l}_{m}\,e^{i\omega t},
     \end{equation}
where $g^{(B)}_{\mu\nu}$ is the background metric (\ref{metric_star})
of a spherically symmetric star
and $\hat{H}$, $\hat{H}_{1},$ and $\hat{K}$ are perturbed metric
functions with respect to $r$. We apply a formalism
developed by Lindblom and Detweiler \cite{Lindblom1985} for
relativistic nonradial stellar oscillations.
The components of the Lagrangian
displacement of fluid perturbations are expanded as
     \begin{eqnarray}
      \xi^{r}      &=& \frac{r^l}{r}e^{\Lambda}\hat{W}Y^{l}_{m}\,
                       e^{i\omega t}, \\
      \xi^{\theta} &=& -\frac{r^l}{r^2}e^{\Lambda}\hat{V}
                       \frac{\partial}{\partial \theta}
                       Y^{l}_{m}\,e^{i\omega t}, \\
      \xi^{\phi}   &=& -\frac{r^l}{r^2\sin^2\theta}e^{\Lambda}\hat{V}
                       \frac{\partial}{\partial \phi}Y^{l}_{m}\,e^{i\omega t},
     \end{eqnarray}
where $\hat{W} $ and $ \hat{V}$ are  functions of $r$.

Assuming that the matter is a perfect fluid and the perturbation is adiabatic,
we have the following perturbation equations derived from Einstein equations:
     \begin{eqnarray}
      &&\frac{d\hat{H}_{1}}{dr}=-\frac{1}{r}
       \left[l+1+\frac{2m}{r}e^{2\Lambda}+4\pi r^2(P-\rho)e^{2\Lambda}\right]
       \hat{H}_{1} \nonumber \\
      &&\hspace{1.2cm}+\frac{1}{r}e^{2\Lambda}
       \left[\hat{H}+\hat{K}+16\pi
(P+\rho)\hat{V}\right],\label{perturbation1} \\
      &&\frac{d\hat{K}}{dr}=\frac{l(l+1)}{2r}\hat{H}_{1}
       +\frac{1}{r}\hat{H}
       -\left(\frac{l+1}{r}-\frac{d\Phi}{dr}\right)\hat{K}
       +\frac{8\pi}{r}(P+\rho)e^{\Lambda}\hat{W}, \label{perturbation2} \\
      &&\frac{d\hat{W}}{dr}=-\frac{l+1}{r}\hat{W}+re^{\Lambda}
       \left[\frac{1}{\gamma P}e^{-\Phi}\hat{X}-\frac{l(l+1)}{r^{2}}\hat{V}
       -\frac{1}{2}\hat{H}-\hat{K}\right], \label{perturbation3} \\
      &&\frac{d\hat{X}}{dr}=-\frac{l}{r}\hat{X}+(P+\rho)e^{\Phi}
       \biggl[\frac{1}{2}\left(\frac{d\Phi}{dr}-\frac{1}{r}\right)\hat{H}
       -\frac{1}{2}\left(\omega^{2}re^{-2\Phi}+\frac{l(l+1)}{2r}\right)
       \hat{H}_{1}  \nonumber \\
      &&\hspace{1.2cm}+\left(\frac{1}{2r}-\frac{3}{2}\frac{d\Phi}{dr}\right)
       \hat{K}-\frac{l(l+1)}{r^{2}}\frac{d\Phi}{dr}\hat{V}
       \nonumber \\
      &&\hspace{1.2cm}-\frac{1}{r}
       \left(\omega^{2}e^{-2\Phi+\Lambda}+4\pi(P+\rho)e^{\Lambda}
-r^{2}\left\{\frac{d}{dr}\left(\frac{1}{r^{2}}e^{-\Lambda}\frac{d\Phi}{dr}
       \right)\right\}\right)\hat{W}\biggr], \label{perturbation4} \\
      &&\biggl[1-\frac{3m}{r}-\frac{l(l+1)}{2}-4\pi r^{2}P\biggr]\hat{H}
       -8\pi r^{2}e^{-\Phi}\hat{X} \nonumber \\
      &&\hspace{1.2cm}+r^{2}e^{-2\Lambda}\left[\omega^{2}e^{-2\Phi}
       -\frac{l(l+1)}{2r}\frac{d\Phi}{dr}\right]\hat{H}_{1}
       \nonumber \\
      &&\hspace{1.2cm}-\left[1+\omega^{2}r^{2}e^{-2\Phi}-\frac{l(l+1)}{2}
       -\left(r-3m-4\pi r^{3}P\right)\frac{d\Phi}{dr}\right]\hat{K}=0,
       \label{perturbation5} \\
      &&\hat{V}=\frac{e^{2\Phi}}{\omega^{2}(P+\rho)}
       \left[e^{-\Phi}\hat{X}
       +\frac{1}{r}\frac{dP}{dr}e^{-\Lambda}\hat{W}+\frac{1}{2}(P+\rho)\hat{H}
       \right] ,  \label{perturbation6}
     \end{eqnarray}
where $\gamma$ is the adiabatic index of the unperturbed stellar
model, which is given by
\begin{equation}
\gamma=\frac{\rho+P}{P}\frac{dP}{d\rho}.
\end{equation}

Furthermore, if the proper boundary conditions are imposed on the
above perturbation equations, the problem to solve becomes an
eigenvalue one.  These boundary conditions are given as follows: (i) the
eigenfunctions are regular at the stellar center, (ii) the Lagrangian
perturbation of pressure vanishes at the stellar surface, and (iii)
the gravitational wave is only an outgoing one at infinity.  The QNMs
of quark stars are determined by solving this eigenvalue problem,
and $\omega$ is the eigenvalue of the above perturbation equations. For
the treatment of the boundary condition at infinity, we adopt the method
of continued-fraction expansion proposed by Leaver \cite{Leaver1985}.
The detailed method of determining QNMs is given
in \cite{Sotani2003,Sotani2001}.

\section{Numerical results}
\label{sec:result}

\subsection{$f$ mode}

As for gravitational wave radiation emitted from a compact star, there
exist QNMs which correspond to each $l$.  In this paper, for
simplicity we discuss such QNMs only for $l=2$, which would be
a dominant mode in a weak field regime.

Compared with the other QNMs, we might relatively easily  detect the $f$
mode because its frequency and damping rate are approximately within
the sensitivity band of  gravitational wave interferometers: e.g.,
LIGO and so on.  Therefore, first we discuss the possibility of
constraining parameters of bag models by using the $f$ mode data. When we
discuss the observational data of frequency Re($\omega$) and damping
rate Im($\omega$) of the $f$ mode, we should bear in mind that the
observational data of frequency will be far more accurate than that of
damping rate.  It was discussed in Ref. \cite{Kokkotas2001} that the
relative error of the frequency will be about three orders of magnitude
smaller than that of the damping rate. To get an insight into the
dependence of the $f$ mode on the bag constant $B$ and the $s$ quark
mass $m_s$, we compute the $f$ mode for each stellar model in Tables
\ref{tab_property_QS1} and \ref{tab_property_QS2}.

\begin{center}
\begin{figure}[htbp]
\includegraphics{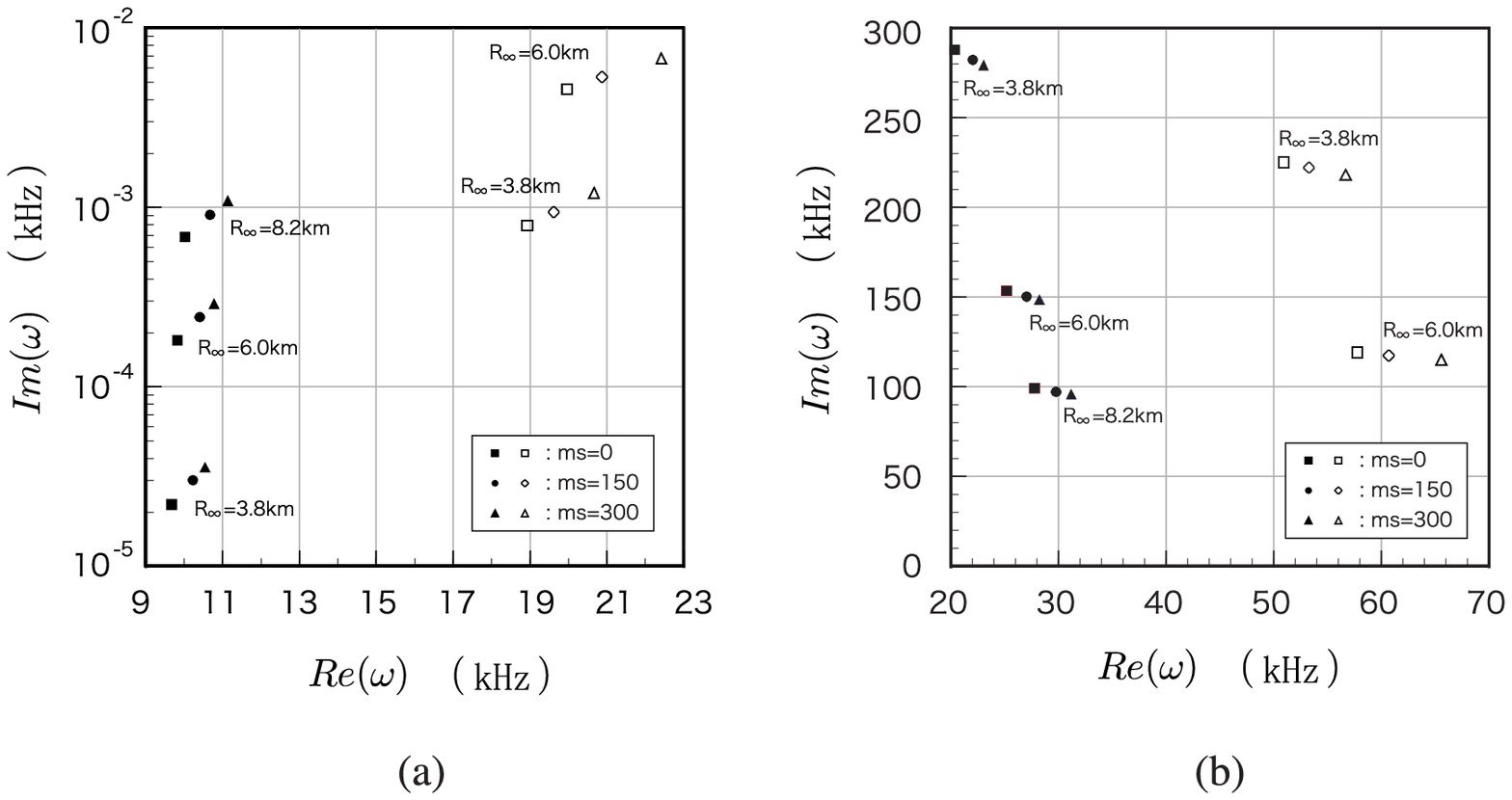}
\caption{
Complex frequencies of (a) $f$ mode and  (b) the lowest $w_{\rm
II}$ mode for $l=2$.  The squares, circles and triangles represent
$m_s=0$, $150$ MeV and $300$ MeV, respectively.  In each panel, the
solid (open) marks  denote the case of $B^{1/4}=145$ MeV ($200$ MeV).
}
\label{fig_fmode}
\end{figure}
\end{center}

In Fig. \ref{fig_fmode}(a) we plot the complex frequencies of $f$ mode
for $l=2$. In the figure, the squares, circles, and triangles represent
$m_s=0$, $150$,  and $300$ MeV, respectively, and  the solid (open)
marks  denote the case of $B^{1/4}=145$ MeV ($200$ MeV).
In Ref. \cite{Sotani2003} the
authors did not consider such $m_s$ dependence because they simply
assumed that $m_s = 0$
as their first step in the paper. From
Fig. \ref{fig_fmode}(a), however,
we find that both the frequencies and
damping rates  change at about  $10\%$ as $m_s$ changes from
zero to $300$ MeV for each $B$, and we see that it is important for us
to include such effects of nonzero values of $m_s$ in bag models to
predict the $f$ mode.

\begin{center}
\begin{figure}[htbp]
\includegraphics{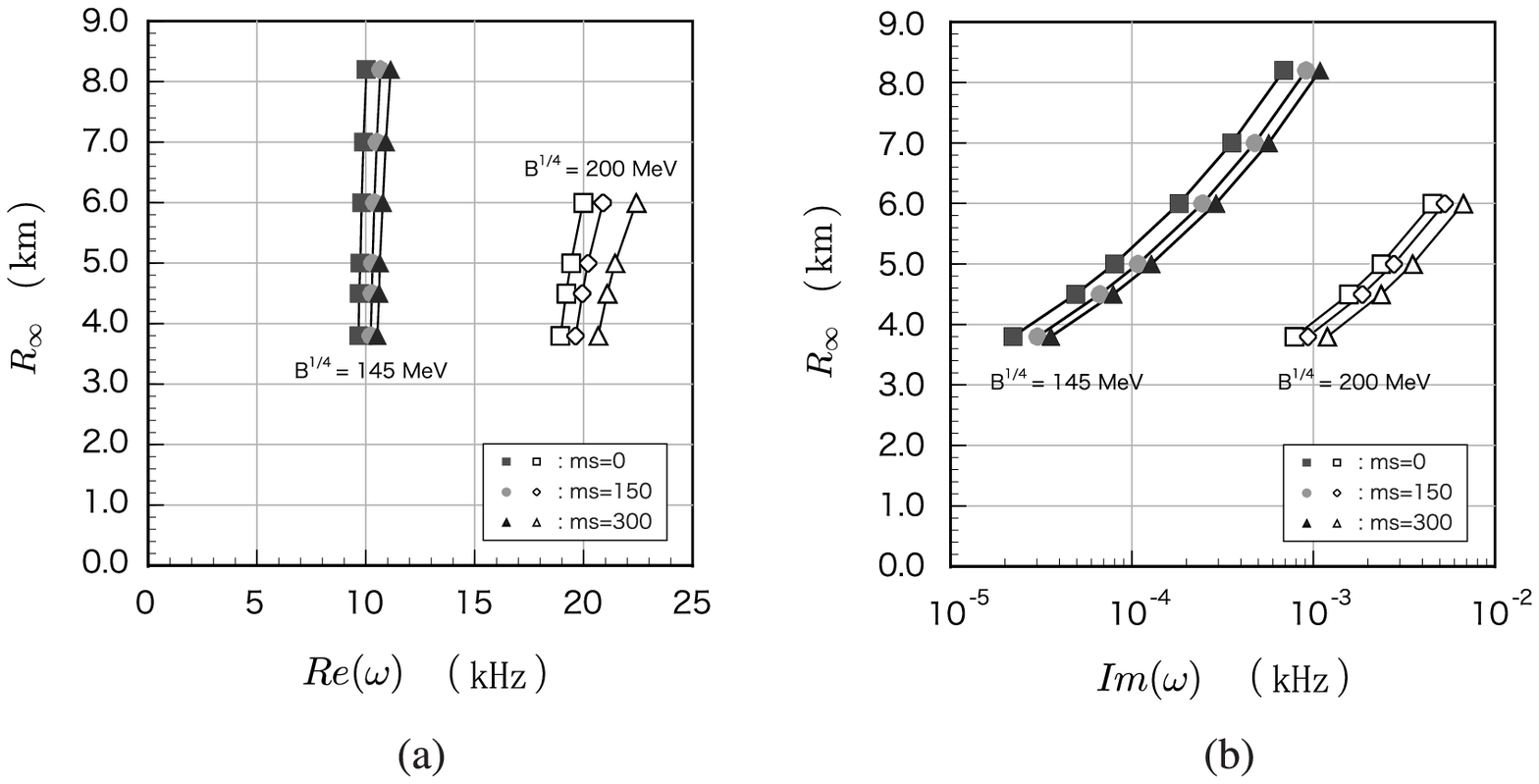}
\caption{
Radiation radius as a function of (a) frequency Re($\omega$) and (b)
damping rate Im($\omega$) of $f$ mode for $l=2$.  In each figure, the
solid and open marks denote $B^{1/4}=145$ MeV and $200$ MeV,
respectively.
}
\label{fig_dependenceRf}
\end{figure}
\end{center}

\begin{center}
\begin{figure}[htbp]
\includegraphics{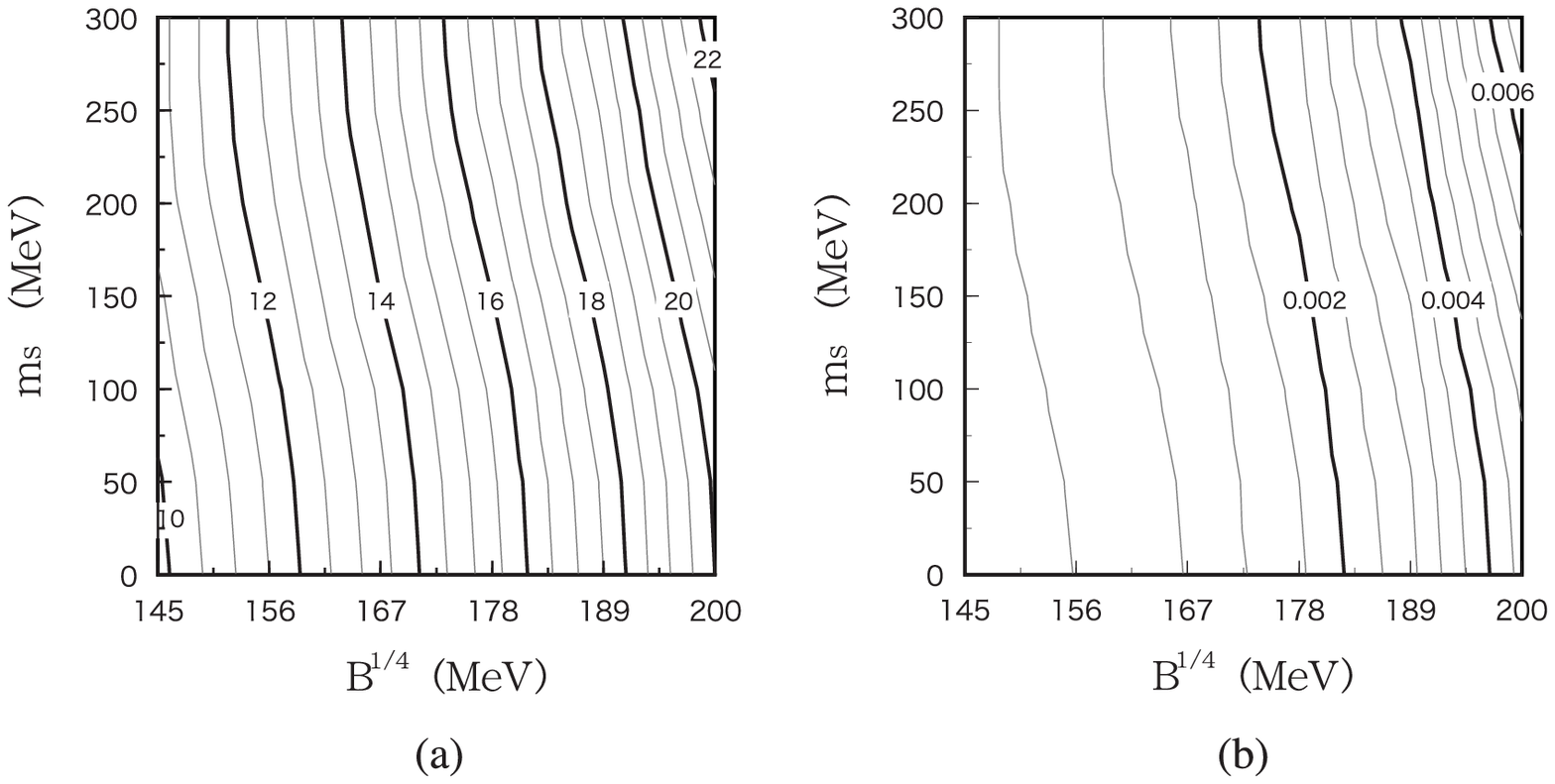}
\caption{
(a) Contours of frequency Re($\omega$) and (b) damping rate
Im($\omega$) of  $f$ mode for $l=2$ in ($B^{1/4}$, $m_s$) plane. Here
we adopt the radiation radius as $R_{\infty}=6.0$ km. The numbers with
the curve denote the value of frequency Re($\omega$) or damping rate
Im($\omega$) in units of kHz.  }
\label{fig_contourf}
\end{figure}
\end{center}

In Fig. \ref{fig_dependenceRf} we plot the radiation radius as a
function of the (a) frequency Re($\omega$) and (b) damping rate
Im($\omega$) of the $f$ mode for $l=2$.  In each panel, the solid and
open marks denote $B^{1/4}=145$ MeV and $200$ MeV, respectively.  From
Fig. \ref{fig_dependenceRf}(b) we see that the damping rate is very
sensitive to the radiation radius, unlike the frequency in
Fig. \ref{fig_dependenceRf}(a). If we  observe the damping rate of
the $f$ mode within $20\%$, the radiation radius can be determined
within $10\%$ as a function of $B$, independently of the value
of $m_s$. On the other hand, from Fig. \ref{fig_dependenceRf}(a) we
see that if we observe the  frequency within about $20\%$,
for example, we can determine the value of $B$ with reasonable accuracy by
using the simultaneously obtained radiation radius from
Fig. \ref{fig_dependenceRf}(b).  To demonstrate that, we investigate
the dependence of the $f$ mode on both the bag constant and the $s$ quark
mass when we fix the  radiation radius as $R_{\infty}=6.0$ km. In
Fig. \ref{fig_contourf}, we plot the contours of the (a) frequency
Re($\omega$) and (b) damping rate Im($\omega$) of the $f$ mode for
$l=2$ in the ($B^{1/4}$, $m_s$) plane.  The numbers on the curves denote
the value of frequency Re($\omega$) or damping rate Im($\omega$) in
units of kHz.  We should note that each contour line is nearly
parallel to the vertical axis in both Figs. \ref{fig_contourf}(a) and
\ref{fig_contourf}(b).
This implies that we can obtain a stringent constraint on the bag
constant. Indeed, Fig. \ref{fig_contourf}(a) shows that if the
frequency of the $f$ mode is detected within $20\%$, then we can
determine the value of the bag constant within about $15\%$
independently of the value of the $s$ quark mass.

\subsection{The lowest $w_{\rm II}$ mode}

In general, there exist infinite number of $w$ modes and several
$w_{\rm II}$ modes for each $l$.
As was clearly shown by numerical analysis in Ref.~\cite{Sotani2003},
however, the $w$ modes are insensitive to the parameters in bag models in
which we are interested here. Therefore, we do not adopt $w$ modes in
this study.
In addition, only one $w_{\rm II}$ mode was found in each quark star
model adopted in this study. That is because the ``compactness''-
i.e., mass to radius ratio---of the adopted quark stars is much
smaller than that of standard neutron stars (see Tables
\ref{tab_property_QS1} and \ref{tab_property_QS2}).  Here, for
simplicity we deal with only the lowest $w_{\rm II}$ mode for $l=2$
\cite{footnote2}.

\begin{center}
\begin{figure}[htbp]
\includegraphics{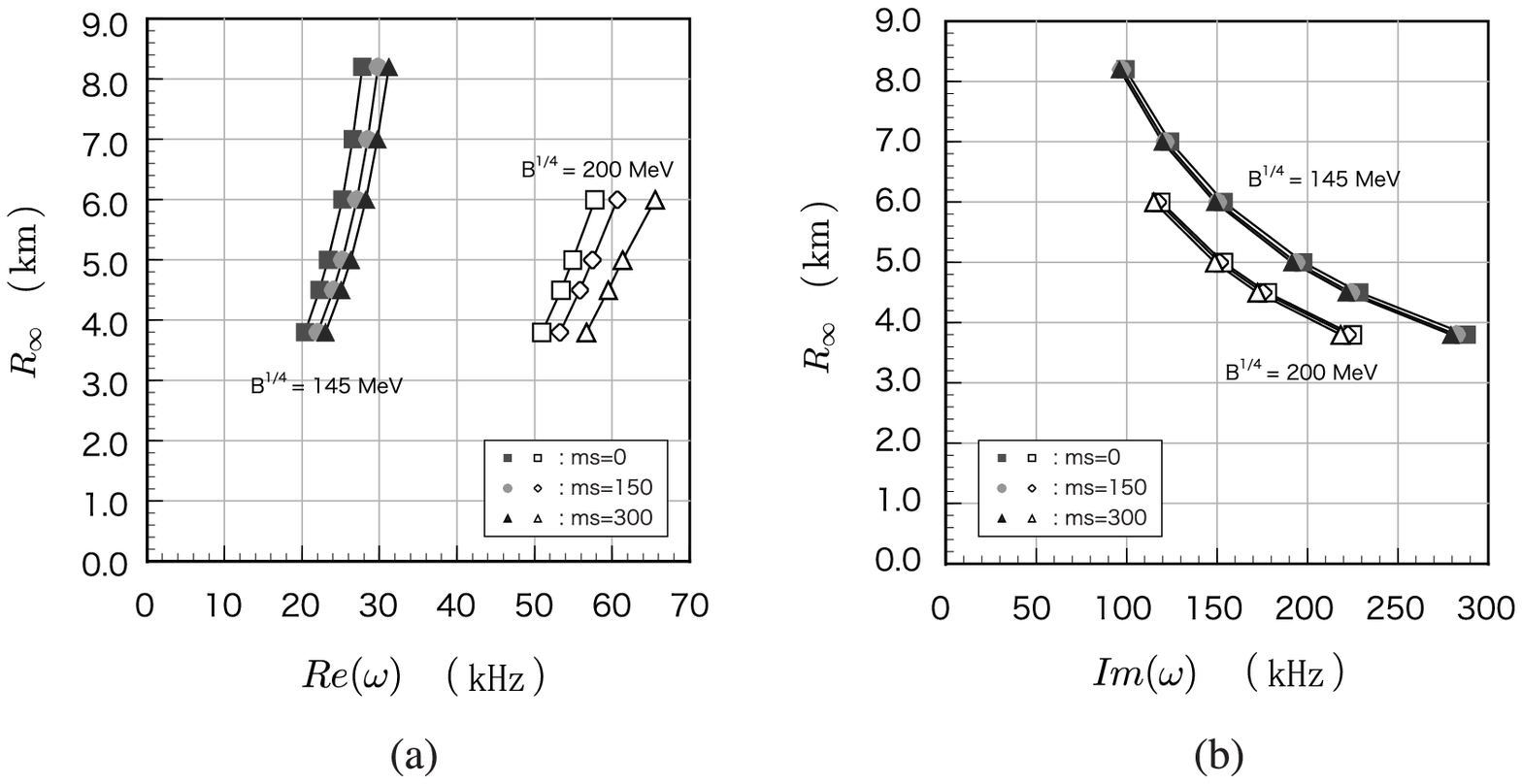}
\caption{
Radiation radius as a function of (a) frequency Re($\omega$) and (b)
damping rate Im($\omega$) of the lowest $w_{\rm II}$ mode for $l=2$.
In each figure, the solid marks and open ones denote
$B^{1/4}=145$ MeV and $200$ MeV, respectively.
}
\label{fig_dependenceRw}
\end{figure}
\end{center}

In Fig. \ref{fig_fmode}(b)  we plot the complex frequencies of the the
lowest $w_{\rm II}$ mode. From Fig. \ref{fig_fmode}(b), we see that
the $m_s$ dependence on the complex frequencies is important, although it
was not considered in Ref. \cite{Sotani2003}. In
Fig. \ref{fig_dependenceRw} we plot the radiation radius as a function
of the (a) frequency Re($\omega$) and (b) damping rate Im($\omega$). The
notes on the marks are the same as those in
Fig. \ref{fig_dependenceRf}. This figure tells us that it has the
same tendency which appears in the case of the $f$ mode; namely, the
damping rate is very sensitive to the radiation radius, unlike the
frequency. Thus, if we directly observe the damping rate, we can
constrain the radiation radius of the quark star.

\begin{center}
\begin{figure}[htbp]
\includegraphics{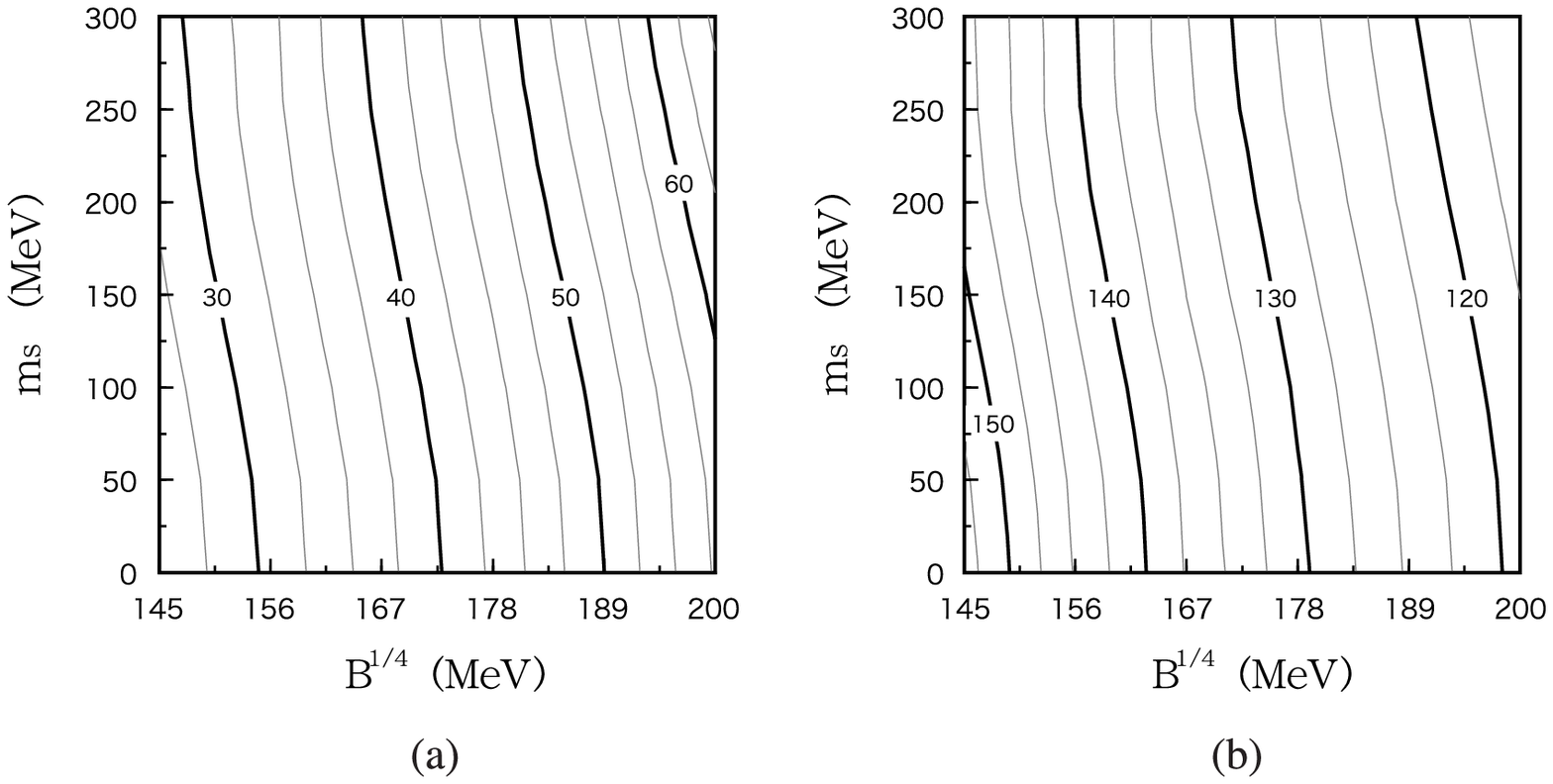}
\caption{
(a) Contours of frequency Re($\omega$) and (b) damping rate
Im($\omega$) of the lowest $w_{\rm II}$ mode for $l=2$ in ($B^{1/4}$,
$m_s$) plane. Here we adopt the radiation radius as $R_{\infty}=6.0$
km. The numbers with the curve denote the value of frequency
Re($\omega$) or damping rate Im($\omega$) in units of kHz.
}
\label{fig_contourw}
\end{figure}
\end{center}

Furthermore, along with the case of the $f$ mode, we plot the contour
of the frequency and the damping rate of the lowest $w_{\rm II}$ mode
in the ($B^{1/4}$, $m_s$) plane in Fig. \ref{fig_contourw}.  By using the
lowest $w_{\rm II}$ mode, we are able to develop a similar argument
as that discussed in the case of the $f$ mode and get an independent constraint on
the bag constant and $s$ quark mass.

However, we should keep in mind that  even if a large amount of energy
is released through these modes, accurate observation of the $w$ and
$w_{\rm II}$ modes may be difficult. That is because both the
frequency and damping rate of these modes are larger than
the sensitivity ranges of gravitational wave
interferometers \cite{Andersson1996}.  Thus we would mainly use the
observational data of the $f$ mode to obtain information about the bag
model parameters. Then we could subsidiarily use the data on the
lowest $w_{\rm II}$ mode. It should be noted that, since
Fig. \ref{fig_contourw} has similar features found in the $f$ mode, even
if we combine it with information from the $f$ mode, we may not have
strict constraints on both $m_s$ and $B$, independently.  There exists
an essential degeneracy in $f$ and the lowest $w_{\rm  II}$ mode QNMs
because the configuration of the contour of  Fig. \ref{fig_contourf}
is very similar to that of Fig. \ref{fig_contourw}.

\section{Conclusion}
\label{sec:conclusion}

We have discussed how we can obtain information about the
EOS of quark matter by using  future observations of gravitational
waves emitted  from quark stars. In particular we have studied the EOS
in bag models and assumed that the star is a pure quark star. We have
computed the QNMs---i.e., the $f$ and the lowest $w_{\rm II}$ modes---in
several quark star models. We have demonstrated that by comparing
the results of theoretical computations with the observational data
of the $f$ mode and the lowest $w_{\rm II}$ modes we can obtain
constraints on the bag constant $B$ and $s$ quark mass $m_{s}$.

If we have the damping rate---i.e., Im($\omega$)---of the $f$ mode within
$20\%$, the radiation radius of the quark star  can be
determined within about $10\%$ including the uncertainty of the
$s$ quark mass.  Furthermore if we  also obtain the frequency---i.e.,
Re($\omega$)---of the $f$ mode within $20\%$, the value of the bag
constant can be determined within about $15\%$, independently of
the uncertainty of the $s$ quark mass, by using the simultaneously
obtained radiation radius.  Concerning the lowest $w_{\rm II}$ mode,
we can also develop a similar argument as in the case of the $f$ mode and get
independent constraints on the model parameters. However, note that it
is relatively difficult to detect the lowest $w_{\rm II}$ mode by the
future planned gravitational wave interferometers whose frequency
ranges are not very sensitive to the lowest $w_{\rm II}$
mode. Therefore, such data will be subsidiarily used in statistical
analyses.  It should be also noted that there is a degeneracy in the
dependence of $f$ and the lowest $w_{\rm II}$ mode QNM complex
frequencies on the bag model parameters $m_s$ and $B$.
As for high frequency gravitational waves,
a dual-type detector has been proposed, which would
reach very good spectral strain
sensitivities ($\sim 2\times 10^{-23}\ \mbox{Hz}$)
in a considerable broadband (between $1$ and
$3\ \mbox{kHz}$)~\cite{Cerdonio2001}
and then open a new interesting window to the
QNMs of compact objects.

As long as  we have data with tolerable accuracy in future, we
will be able to perform statistical analyses to fit them and get
constraints on both the bag constant and the $s$ quark mass. If we
have further information about  the bag constant or the $s$ quark mass
by utilizing the EOS models based on future developments in an
effective theory of QCD, such as perturbation theories or
finite-temperature lattice data, we can constrain them more
strictly. Including independent observations of the radiation radius---e.g.,
sorts of X-ray observations---would also support us in inferring the
bag model parameters. QNMs from compact stars will be observed in the near
future and deepen our understanding of hadron physics and QCD.

\acknowledgments

We would like to thank K. Maeda for useful discussions. K.K. was
supported by the Grants-in Aid of the Ministry of Education, Science,
Sports, and Culture of Japan (Grant No.15-03605).  TH was supported by the JSPS.



\begin{table}[htbp]
      \begin{center}
          \leavevmode
          \caption{
          Bag model parameters fitted to hadron mass spectra.
          }
          \begin{tabular}{cccccc}
            \hline\hline
              & $B^{1/4}$ (MeV) & $m_{s}$ (MeV)& $\alpha_{c}$ &
              Reference& \\ \hline
              & 145
              &  279
              & 2.2
              & \cite{DeGrand:cf}
              &\\
              & 200--220
              & 288
              & 0.8--0.9
              & \cite{Carlson:er}
              &\\
              & 149
              & 283
              & 2.0
              & \cite{Bartelski:1983cc}
              &\\
            \hline\hline
          \end{tabular}
          \label{table:fitting}
      \end{center}
\end{table}

\begin{table}[htbp]
\caption{
Properties of quark stars in the case of $B^{1/4}=145\,\,\text{MeV}$.
Here, $R_{\infty}$, $m_s$, $\rho_c$, $M/M_{\odot}$, $R$, and $M/R$
are  the radiation radius, the $s$ quark mass,
the central density, the mass in units of $M_{\odot}$,
the radius, and the compactness
of the star, respectively.
}
\label{tab_property_QS1}
     \begin{center}
       \begin{tabular}{cccccc}
            \hline\hline
         $R_{\infty}\,\,(\text{km})$ &
         $m_s\,\,(\text{MeV})$  &  $\rho_{c}\,\,(\text{g/cm}^3)$  &
         $M/M_{\odot}$ & $R\,\,(\text{km})$  & $M/R$\\  \hline
           $3.800$ & $0.0$  &  $4.221\times 10^{14}$
                       &  $4.539\times 10^{-2}$ & $3.731$ &
$1.796\times 10^{-2}$   \\
           $3.800$  & $150.0$ & $4.714\times 10^{14}$
                       &  $5.020\times 10^{-2}$  & $3.724$ &
$1.991\times 10^{-2}$   \\
           $3.800$   & $300.0$ & $5.027\times 10^{14}$
                       &  $5.305\times 10^{-2}$ & $3.719$   &
$2.106\times 10^{-2}$   \\
           $6.000$   & $0.0$ & $4.412\times 10^{14}$
                       &  $1.678\times 10^{-1}$ & $5.735$ &
$4.320\times 10^{-2}$   \\
           $6.000$   & $150.0$ & $4.979\times 10^{14}$
                       &  $1.846\times 10^{-1}$ & $5.706$ &
$4.778\times 10^{-2}$   \\
           $6.000$   & $300.0$ & $5.392\times 10^{14}$
                       &  $1.951\times 10^{-1}$ & $5.688$ &
$5.066\times 10^{-2}$   \\
           $8.200$   & $0.0$ & $4.731\times 10^{14}$
                       &  $3.925\times 10^{-1}$ & $7.544$ &
$7.684\times 10^{-2}$   \\
           $8.200$   & $150.0$ & $5.436\times 10^{14}$
                       &  $4.293\times 10^{-1}$ & $7.472$ &
$8.483\times 10^{-2}$   \\
           $8.200$   & $300.0$ & $6.049\times 10^{14}$
                       &  $4.540\times 10^{-1}$ & $7.423$ &
$9.031\times 10^{-2}$  \\
             \hline\hline
      \end{tabular}
     \end{center}
\end{table}

\begin{table}[htbp]
\caption{Same as Table \ref{tab_property_QS1}
except for $B^{1/4}=200\,\, \text{MeV}$.}
\label{tab_property_QS2}
     \begin{center}
       \begin{tabular}{cccccc}
             \hline\hline
         $R_{\infty}\,\,(\text{km})$ &
         $m_s\,\,(\text{MeV})$  &  $\rho_{c}\,\,(\text{g/cm}^3)$  &
         $M/M_{\odot}$ & $R\,\,(\text{km})$  & $M/R$\\  \hline
           $3.800$ & $0.0$ & $1.660\times 10^{15}$
                       &  $1.477\times 10^{-1}$ & $3.560$ &
$6.125\times 10^{-2}$ \\
           $3.800$ & $150.0$ & $1.796\times 10^{15}$
                       &  $1.563\times 10^{-1}$ & $3.544$ &
$6.513\times 10^{-2}$ \\
           $3.800$ & $300.0$ & $2.019\times 10^{15}$
                       &  $1.692\times 10^{-1}$ & $3.520$ &
$7.098\times 10^{-2}$ \\
           $6.000$ & $0.0$ & $2.038\times 10^{15}$
                       &  $4.742\times 10^{-1}$ & $5.113$ &
$1.369\times 10^{-1}$ \\
           $6.000$ & $150.0$ & $2.285\times 10^{15}$
                       &  $4.977\times 10^{-1}$ & $5.052$ &
$1.455\times 10^{-1}$ \\
           $6.000$ & $300.0$ & $2.781\times 10^{15}$
                       &  $5.344\times 10^{-1}$ & $4.953$ &
         $1.593\times 10^{-1}$  \\
             \hline\hline
        \end{tabular}
     \end{center}
\end{table}

\end{document}